\documentclass[fleqn,usenatbib]{mnras}

\usepackage[T1]{fontenc}
\usepackage{ae,aecompl}

\usepackage{graphicx}	
\usepackage{amsmath}	
\usepackage{amssymb}	
\usepackage{color,ulem}
\usepackage[dvipsnames]{xcolor}
\usepackage[T1]{fontenc}
\usepackage[utf8]{inputenc}
\usepackage{hyperref}

\usepackage{txfonts}

\newcommand{\rdin}{r_{{\rm ej}}^{{\rm in}}}
\newcommand{\rdout}{r_{{\rm ej}}^{{\rm out}}}
\newcommand{\vdout}{\beta_{{\rm ej}}^{{\rm out}}}
\newcommand{\rext}{r_{{\rm t}}}
\newcommand{\vext}{v_{{\rm t}}}

\newcommand{\rin}{r_{{\rm in}}}
\newcommand{\rout}{r_{{\rm out}}}

\definecolor{electricpurple}{rgb}{0.75, 0.0, 1.0}
\definecolor{blazeorange}{rgb}{1.0, 0.4, 0.0}
\definecolor{st.patrickblue}{rgb}{0.14, 0.16, 0.48}
\definecolor{sacramentostategreen}{rgb}{0.0, 0.34, 0.25}

\DeclareSymbolFont{cmletters}{OML}{cmm}{m}{it}
\DeclareMathSymbol{v}{\mathalpha}{cmletters}{"76}
\title[Jet breakout from NS merger remnant]{The $\gamma$-rays that accompanied GW170817 and the observational signature of a magnetic jet breaking out of NS merger ejecta.}

\author[O. Bromberg, A. Tchekhovskoy, O. Gottlieb, E. Nakar, T. Piran]{O. Bromberg,$^{1}$\thanks{E-mail:omer@wise.tau.ac.il} A. Tchekhovskoy,$^{2}$ O. Gottlieb,$^1$ E. Nakar,$^{1}$ and T. Piran$^{3}$
\\
$^{1}$ The Raymond and Beverly Sackler School of Physics and Astronomy,
Tel Aviv University, Tel Aviv 69978, Israel\\
$^{2}$Center for Interdisciplinary Exploration \& Research in Astrophysics (CIERA),
Physics \& Astronomy, Northwestern University, Evanston, IL 60202, USA\\
$^{3}$Racah Institute of Physics, The Hebrew University of Jerusalem,
Jerusalem 91904, Israel\\
}
\date{Accepted XXX. Received YYY; in original form ZZZ}

\pubyear{2015}

\begin{document}
\label{firstpage}
\pagerange{\pageref{firstpage}--\pageref{lastpage}}
\maketitle

\begin{abstract}
We present the first relativistic MHD numerical simulation of a magnetic jet that propagates through and emerges from the dynamical ejecta of a binary neutron star merger. 
Generated by the magnetized rotation of the merger remnant, the jet propagates through the ejecta 
and produces an energetic  cocoon that expands at mildly relativistic velocities and breaks out of the ejecta.
 We show that if the ejecta has a  low-mass ($\sim10^{-7} M_\odot$) high-velocity  ($v\sim0.85$ c) tail, the cocoon shock breakout will generate $\gamma$-ray emission  that is comparable to  the observed short GRB170817A that accompanied  the recent gravitational wave  event  GW170817. Thus, we propose that this GRB, which is quite different from all other short GRBs observed before, was produced by a different mechanism.  We expect, however, that such events are numerous and many will be detected in coming LIGO-Virgo runs.

\end{abstract}

\begin{keywords}
keyword1 -- keyword2 -- keyword3
\end{keywords}



\defcitealias{2016MNRAS.456.1739B}{BT16}

\section{Introduction}

Merging  neutron stars (NS) are considered to be natural candidates for both gravitational waves (GW) events and short hard gamma-ray bursts \citep[sGRBs;][]{Eichler1989}. However while the $\gamma$-ray emission is beamed at an angle $\lesssim20^\circ$ from the rotational axis 
\citep{nakar_short_grbs_2007,2015ApJ...815..102F}, the GW signal has a wide angle distribution.
Given the rather small opening angle of the $\gamma$-ray emission and
the expected rate of both events, it is unlikely that the two will be detected simultaneously from the same source. 
It is therefore, quite interesting that the first detection of a GW signal from a merger \citep{2017GCN.21505....1S,2017GCN.21513....1S,2017GCN.21527....1S,LIGO} was accompanied by a $\gamma$-ray signal \citep{Multi,2017GCN.21528....1S}, suggesting that the $\gamma$-ray photons are emitted on a much wider angle than just the jet angle. 

Wide angle emission of soft $\gamma$-rays is seen also in a fraction of long GRBs which originate from collapsing massive stars \citep{1998Natur.395..663K,2006Natur.442.1008C,sod_subenergetic_06,2007ApJ...654..385K}, knowns as low-luminosity GRBs ({\it ll}GRBs). In this situation the collapse of the stellar core leads to the formation of a relativistic jet, similar to jets of sGRBs. The jet propagates through the stellar envelope forming an energetic cocoon that contains the energy injected by the jet pushing the stellar material sideways \citep[e.g.][]{mac99,2007ApJ...665..569M,2011ApJ...740..100B,2013ApJ...767...19L,2016MNRAS.456.1739B,2017arXiv170706234H}. If the jet emerges from the stellar envelope it produces a regular highly beamed long GRB. 
If the jet fails while inside the star only the cocoon will still break out, leading to a {\it ll}GRB with a relatively soft spectrum and a wide angle emission \citep{2012ApJ...747...88N,2015ApJ...807..172N}. Interestingly, direct observational evidence based on the duration distribution of long GRBs support this picture \citep{2012ApJ...749..110B} and suggests that the number of choked jets producing {\it ll}GRBs is much larger than successful ones. 

In NS mergers, the coalescence process involves  ejection of $\sim10^{-3} - \text{few}\times10^{-2}M_\odot$ of dynamical mass ejecta (DME) due to tidal interactions of the two stars \citep{Davies1994,Ruffert1996,Rosswog1999,Bauswein2013,Hotokezaka2013,Sekiguchi2015,Radice2016}. Once launched, the sGRB jet must propagate through this ejecta, which plays a similar role of stellar envelopes in the case of long GRBs. 
\citep{Nagakura2014,Murguia2014,Duffell2015,nakar2017,g17,Gottlieb17b}.  An important difference is that the jet propagates in matter that  is moving at sub-relativistic velocities ($\sim 0.1\text{-}0.4$ c) while a stellar ejecta is static. 
Inspection of the duration distribution of sGRBs supports this picture, by revealing evidence that these events are indeed surrounded by a few percent of solar mass through which their jets must penetrate \citep{Moharana2017}.

A second component of ejecta arises in mergers from neutrino driven winds. 
If the merger product is a rapidly rotating (a few ms) massive proto-NS (PNS) it  has an initial radius of $\sim30$ km and temperatures of $\gtrsim10$ Mev \citep{2011PhRvL.107e1102S,2017ApJ...846..114F}. It is cooled via massive neutrino emission, which drives a strong baryonic wind from the surface of the PNS with power of $\lesssim10^{50}$ erg/s and mass flux of $\lesssim10^{-3} M_\odot$/s  \citep{2017ApJ...846..114F}. The wind maintains its power over a time scale of $\text{few} \times 100$~ms to $\sim1$~s \citep{2011PhRvL.107e1102S, 2017ApJ...846..114F}. The wind loads the magnetosphere around the PNS with heavy material and imposes hydrodynamic conditions. Once the PNS cools down it contracts to a regular NS size ($\sim10$ km) and the wind relaxes. Neutrino driven wind can also arise from a disk  that forms regardless whether the 
compact object is a neutron star or a black hole. {Combined, the DME and the neutrino driven wind are usually referred to as the ``ejecta" surrounding the merger \citep[see e.g.][for a detailed discussion of the different ejecta components]{hotokezaka2015}.}

Magnetic fields are likely to play an important role in the  evolution of the system. 
Various amplification processes during the merger \citep{2006Sci...312..719P, 2013ApJ...769L..29Z,2015PhRvD..92l4034K} or within the newly formed PNS \citep{2017MNRAS.471.1879G} can increase the poloidal magnetic field on the surface of the PNS component to an order of $>10^{14}$~G during the initial phase. Conservation of angular momentum and magnetic flux during the contraction phase increases the field by another order of magnitude and spins up the PNS. The rapid rotation coils up the magnetic fields and converts PNS rotational energy into an outflowing electromagnetic Poynting flux and a pair of powerful magnetically dominated jets. In this {protomagnetar model} \citep{2011MNRAS.413.2031M} the jets can easily power the observed $\gamma$-ray emission in sGRBs. In the case of an accreting BH a similar situation takes place, where as magnetic field is continuously supplied by the accretion disk, in which  magnetic amplification processes can occur \citep{tch11,tch10b}.
This motivates the study of the jets and their interaction with the ejecta in the context of magnetohydrodynamics (MHD) rather than hydrodynamics.  

Here, we provide the first 
numerical study of the interaction of relativistic Poynting flux dominated jets with a characteristic ejecta {that is expands sub-relativistically, focusing on } the context of sGRBs. We study the properties of the jet and its cocoon during the propagation through the ejecta and follow them throughout the breakout and the expansion into the surrounding medium.  We show  that the cocoon contains enough energy to produce a strong emission over wide angles once it exits the ejecta. We then show that if the ejecta is surrounded by a tail of low mass ($\sim10^{-7} M_\odot$) fast material ($v\sim0.85$ c), the breakout can generate $\gamma$-ray emission  detectable out to distances of  $\lesssim100$~Mpc, similar to the distance over which GWs are detected from the merger. Thus the cocoon breakout emission can be seen in coincidence with the GW event. 
We compare our results with the observed high energy emission, GRB170817A
\citep{2017GCN.21528....1S,Multi},
from the recent gravitational wave (GW) event, GW170817 \citep{2017GCN.21505....1S,2017GCN.21513....1S,2017GCN.21527....1S,LIGO}, and show qualitatively that such a scenario can explain the observed emission. A quantitative fit of the breakout emission to the observations is provided in a companion paper that discusses hydrodynamic sGRB jets \citep{Gottlieb17b}.

We use a rotating dipole as a boundary condition for the magnetic jet, corresponding to a PNS. However,  the details of the jet propagation at large radii are not sensitive to the jet launching process and thus our results are appropriate both for jets driven by a PNS or by an accreting black hole. 
For simplicity, throughout the paper we refer to the jet engine as PNS. 
{We begin in \S \ref{sec:simulation} with a discussion of the setup of the numerical simulation and the initial conditions.   We continue with a discussion of the evolution of the system in \S \ref{sec:evolution}. In \S \ref{sec:radiation} we discuss the post-processing of the MHD simulations for calculation of the observed radiation from the system. We compare the results, in \S \ref{sec:compare} with the observations of GRB170817A that accompanied GW170817. We summarize the results in \ref{sec:summary}.}

\section{Simulation setup}
\label{sec:simulation}
To study the jet propagation we run relativistic MHD simulations using HARM, a general relativistic MHD code \citep{gam03,nob06,tch07,mb09,tch11}. The code uses modified spherical polar coordinates ($r$, $\theta$) that span the range $(\rin,\rout)\times(0,\pi)$. To resolve the jet and the cocoon at long distances we moderately concentrate the grid cells toward the polar axis, by deforming the radial grid lines into parabolas (see \citetalias{2016MNRAS.456.1739B}, hereafter \citealt{2016MNRAS.456.1739B}). 
We treat the inner boundary as a perfectly conducting sphere threaded with dipolar magnetic field lines and will refer to it as the star. To generate the jet, at $t = 0$ we spin up the star around the $z$-axis to an angular frequency $\Omega=\tilde{\Omega}c/\rin$. The rotation coils up the field lines creating an azimuthal component $B_\varphi$ and an electric field $E_\theta$. 
The cross product of the two gives a radial Poynting flux 
\begin{equation}\label{eq:LEM}
L_{\rm EM}=\int\frac{c}{4\pi}\left({\bf E}\times{\bf B}\right)_rr^2d\Omega\simeq \frac{1}{4c}B_p^2\Omega_{\rm NS}^2 r_{\rm NS}^4\theta_{\rm open}^4,
\end{equation}
where $B_{\phi}\simeq E_\theta=-B_{r}{\Omega_{\rm NS}r_{\rm NS}\sin\theta}/{c}$.
The rotation creates a ``light cylinder'' at a radius $R_{\rm L}=c/\Omega=\rin/\tilde{\Omega}$ beyond which the flow becomes relativistic. Field lines that cross this radius can no longer co-rotate with the central object and open up. Those open field lines carry the Poynting flux into the jet. In the dipole configuration $\theta_{\rm open}\simeq\sqrt{r/R_L}=\sqrt{\Omega r/c}$, giving the well known dipole formula for the dipole energy losses. 

Since all the important physics occurs at radii $r>R_{\rm L}$, the inner boundary must be placed inside the light cylinder. To maximize the time step (and hence speed) of the simulation we set $\tilde\Omega = 0.8$, or $\rin=0.8 R_{\rm L}$, place the inner ejecta radius at $\rdin=2.5 R_{\rm L}$, and the outer boundary radius at $\rout=10^4 \rin$. We use a resolution of $1024\times432$ cells in the $r$- and $\theta$-directions respectively. At this resolution grid cells maintain an aspect ratio $<3$ throughout the grid, the jet is resolved by $50$ cells across its half-opening angle and the cocoon by additional $90$ cells. 

\begin{figure}
	\includegraphics[width=\columnwidth]{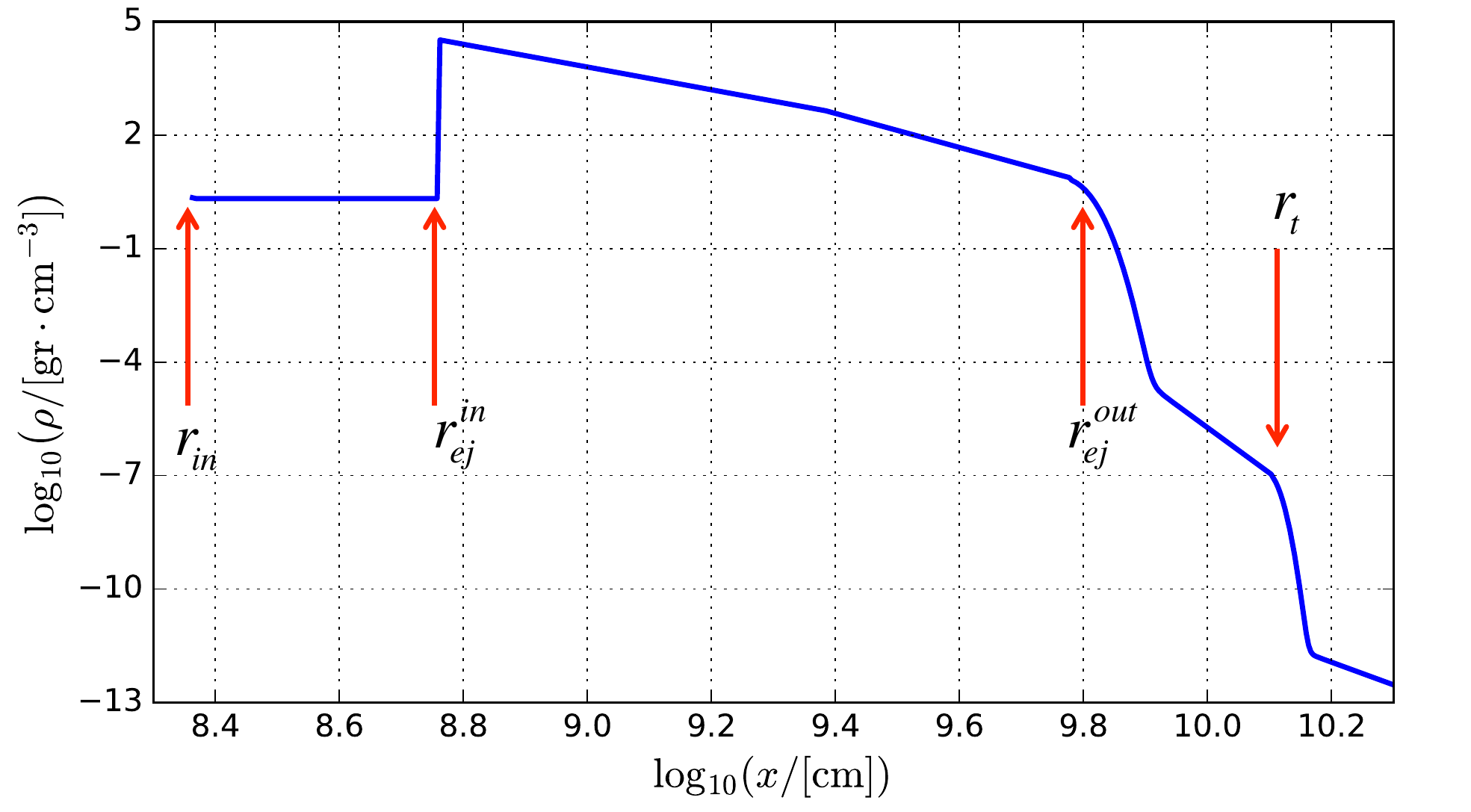}
    \caption{The initial mass distribution around the PNS at $t=0.5$~s after the merger, when we start the simulation. By this time ejecta has expanded to $\rdout=6\times10^9$ cm. It is surrounded by a fast, low density tail that extends the density to $\rext=1.2\times10^{10}$ cm. The gap between $\rin$ and $\rdin$ is filled with matter from the neutrino driven wind, which preceded the jet launching.}
    \label{fig:boundary}
\end{figure}

Our setup approximates the conditions around a PNS  half a second after its creation. By this time, the neutrino driven wind from the surface has sufficiently relaxed and the magnetization in the vicinity of the PNS increased, thus a relativistic Poynting flux dominated jet can be formed.   The ejecta at this time has expanded out to a radius of $\sim10^{9}$~cm from the central object, and a low density region around the PNS forms, filled with the baryonic wind material (see Fig.~\ref{fig:boundary}).
In a black hole disk system such a delay may arise due to the time it takes to the massive PNS to collapse. 
We stress that while this setup is based on a PNS  the question of the jet propagation  and the cocoon that forms that interest us doesn't depends critically on how the jet is launched. Therefore, our results concerning the cocoon shock breakout and the corresponding $\gamma$-rays is applicable to different central engines that launch MHD jets. 

We model the ejecta using the mass density and velocity profiles from \citet{2013PhRvD..87b4001H}, where we slightly modified the velocity profile to make it spherically symmetric. With the total mass of $\sim0.06 M_\odot$, the ejecta extends from $\rdin\simeq 6\times10^8$ cm to $\rdout\simeq6\times10^{9}$ cm and features a homologous relativistic 4-velocity profile, $u(r)=0.4c\times(r/\rdout)$, where $v=uc/\Gamma$ is the 3-velocity and $\Gamma = (1-v^2/c^2)^{-1/2}$ is the Lorentz factor. The ejecta is surrounded by a light tail of faster material, with a mass of $M_{\rm ext}=10^{-7} M_\odot$ and a density profile $\rho(r)\propto r^{-12}$. It expands homologously as well, with a peak velocity of $\vext=0.85c$. We were motivated to add this tail from observational indications suggesting the existence of low mass, high velocity material in the system that created the event GW170817 \citep{mansi}. 
A fast tail of the dynamical ejecta with $v_{ej}\gtrsim 0.6c$ is likely to arise
from the joint interface of the merging neutron stars when the shock breaks out from the surface of the merging objects \citep{Kyutoku2014}. Although it is hard to resolve numerically a small amount of fast moving components, some numerical simulations  suggest that such a fast tail exists \citep{Bauswein2013, Hotokezaka2013},
and that it can contain as much as $\sim 10^{-5}M_{\odot}$. To avoid confusion we refer to the ejecta's core and to its fast, low density tail separately. We use the term ``{\it core ejecta}'' or just ``{\it ejecta}'' for the first, and the term {\it ``tail''} for the later.
The tail does not affect the dynamics of the jet due to its low mass, however it {determines the conditions during the breakout of the cocoon driven shock that produces the observed emission (see Sec.~\ref{sec:radiation}).}
We populate the interior of the ejecta, $r<\rdin$, with a hot low-density medium, that was ejected from the PNS surface by the neutrino driven wind in the first $0.5$~s after the collapse. We assume a wind mass flux of $10^{-3} M_\odot\,{\rm s}^{-1}$ and an energy flux of $10^{50}$ erg/s. The initial profile is illustrated in Fig.~\ref{fig:boundary}. 

\section{The evolution of the system}
\label{sec:evolution}

\subsection{Jet launching and propagation}\label{sec:launch}
As we discussed in Sec.~\ref{sec:launch}, the rotation of the star
launches an electromagnetic wind of luminosity $L_{\rm EM}$ (eq.~\ref{eq:LEM}). We set the magnetic field strength at $\rin$ so that $L_{\rm EM}=4\times10^{50}$ erg/s for a one-sided jet. 
This is equivalent of having a millisecond PNS with a surface poloidal field of $\sim4\times10^{15}$ G and a radius of $12$ km. 
After $2$~s, we shut off the energy injection into the jet by instantaneously stopping the stellar rotation.

\begin{figure}
	\includegraphics[width=\columnwidth]{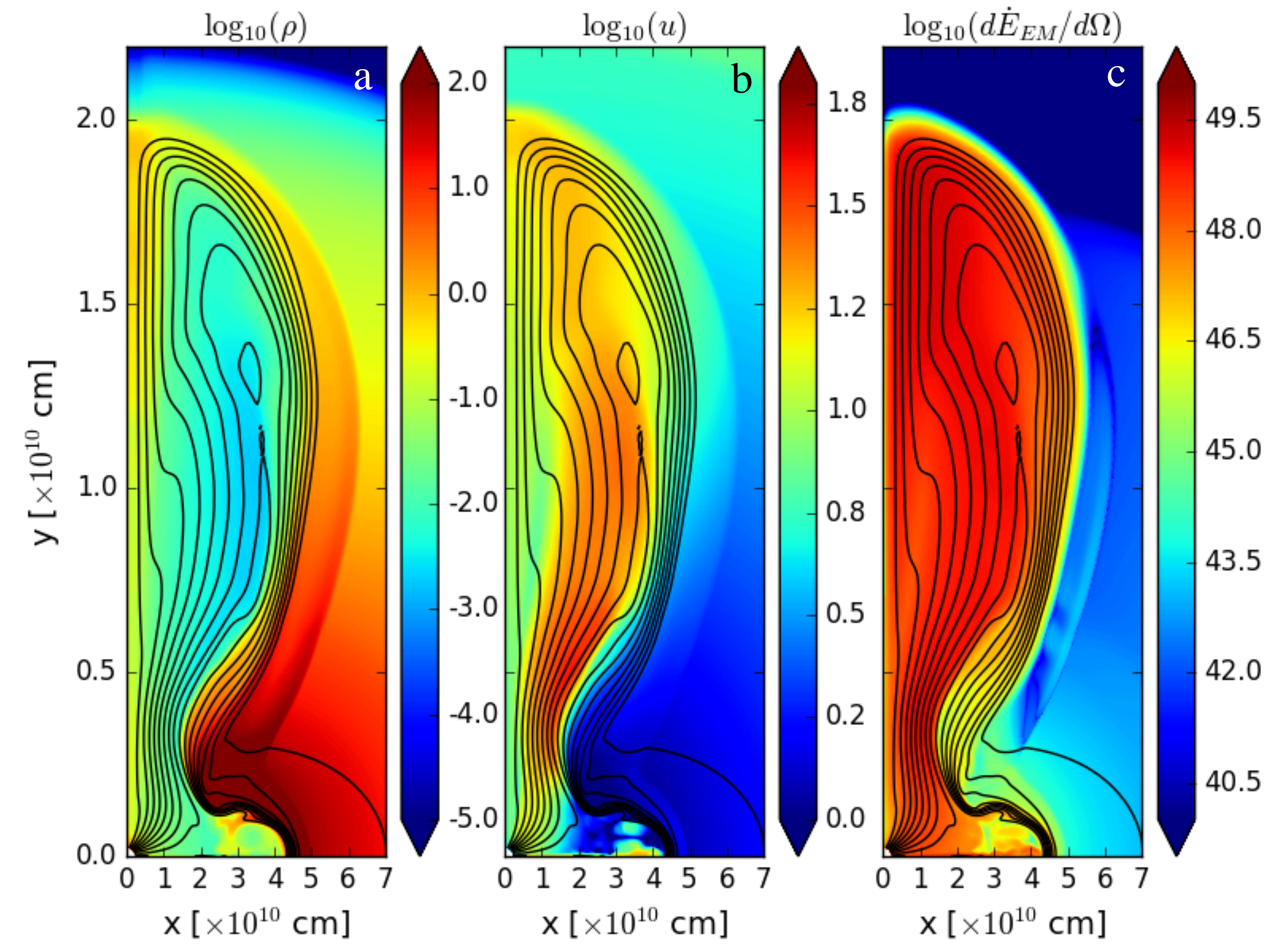}
    \caption{Vertical slices through the logarithms of density (left), 4-velocity (middle) and EM energy flux per unit solid angle (right) at $t=1$~s after jet launching. Red colour shows high and blue low values (see colour bars).}
    \label{fig:jet_core}
\end{figure}

The toroidal field expands nearly radially at essentially the speed of light.
As it encounters the ejecta at $r>\rdin$, it slows down, and toroidal pressure builds up. The toroidal field tension results in 
hoop stress that collimates the field lines towards the rotational axis \citep{2012MNRAS.427.1497L}. 
The Poynting flux, now focused into a small opening angle, eventually attains enough pressure to drill through the ejecta and leads to a jet engulfed by a cocoon of hot shocked gas, as seen in Fig.~\ref{fig:jet_core}. This is very similar to a magnetic jet propagating through a stationary medium \citep{2016MNRAS.456.1739B}. The energy, expended by the jet in drilling through the medium, goes into the cocoon and is stored there in the form of pressure.

The jet injects energy into the cocoon at a rate of $L_j(1-\beta_j)$ where $\beta_j=v_j/c$ is the propagation velocity of the jet's head.
As long as the jet propagates in the core ejecta it maintains sub-to-trans relativistic velocity (with average velocity of $\beta_j\simeq0.7$). 
Once it emerges from the ejecta into the low density tail it becomes highly relativistic and the energy injection stops. 
The total energy of the cocoon can thus be estimated by 
\begin{equation}\label{eq:Ec}
E_{\rm c}\simeq L_jt_{\rm ej}\left(1-\beta_j\right)=L_j\frac{\rdout}{c}\frac{1-\beta_j}{\beta_j-\vdout}\simeq1.3\times10^{50}~{\rm ergs},
\end{equation}
where $t_{\rm ej}=1$ sec is the time the jet emerges from the core ejecta (see below), $\rdout$ is the outer radius of the core ejecta at the time jet is launched, and $\vdout=0.4$ is the velocity of the ejecta's edge. 

Figure~\ref{fig:jet_core} shows the jet just before it emerges from the core ejecta, Panel (a) shows the  
$\log_{10}\rho$, panel (b) 
shows the logarithm of 4-velocity and panel (c) shows the 
logarithm of the EM Poynting flux per unit solid angle. 
The black contours track the poloidal field lines that are anchored to the inner boundary. The field lines extend out to the jet head and then turn around to return back to the equator where they meet with their counterparts from the opposite jet (not shown here). The surface of the jet is the ``null poloidal magnetic field surface'' at which the poloidal magnetic field changes its sign and therefore vanishes. Fig.~\ref{fig:jet_core}c shows that that almost all of the EM flux of the jet is carried out along the outgoing field lines, with the returning field lines carrying a negligible amount of flux. 
The jet is pinched both at the point where it is first collimated by the ambient medium and close to its head. At both of these locations magnetic dissipation is expected to take place and heat up the jet material. \citet{2016MNRAS.456.1739B} showed that magnetic dissipation driven by 3D kink modes, takes place at the collimation point and dissipates the field to a state of equipartition between the thermal and magnetic energies. Because our 2D simulations do not account for such intrinsically 3D dissipation, the jets here are artificially strongly magnetized and cold. The jet is surrounded by a pressurized cocoon which resides between the null poloidal surface and the outer, bow shock (seen in fig.~\ref{fig:jet_core}a as a sharp jump in the density). It is composed of an inner light, magnetized part of the returning field lines and an outer heavier, unmagnetized shocked ejecta part. 
Similar structure of an inner and outer part is seen also in cocoons of hydrodynamic jets \citep[e.g.][]{2011ApJ...740..100B,g17,2017arXiv170706234H}. The main difference is that in magnetized cocoons, magnetic tension inhibits the mixing between the two parts. This effect is more extreme in 2D simulations, which show almost no such mixing. 
 

\subsection{Jet breakout}\label{sec:jet}
The jet breaks out of the core ejecta at time $t_{\rm ej}\simeq1$~sec after launching (see fig. \ref{fig:jet_core}). The corresponding minimal engine activity time that results in a jet breakout is 
\begin{equation}
  T_{min}= t_{\rm ej}\times(1-\beta_j)=0.3~{\rm s},
\end{equation}
in agreement with predictions from short GRB observations \citep{2017MNRAS.472L..55M}. 
The diluted gas in the outer fast tail is too light to hinder the jet propagation, and the jet accelerates to high Lorentz factors immediately after it exits the ejecta. It assumes a conical shape with an opening angle of $\sim20^\circ$. At $t=2$~s the rotation of the inner boundary is stopped and energy injection into the jet shuts off. At this point the jet head is located at $\gtrsim 5\times 10^{10}$ cm. Since the head is highly relativistic by now, the surface of last energy injection cannot catch up right away. It trails the head by a distance of $\sim 5\times 10^{10}$~cm throughout the simulation. The region between the jet head and the surface of the last energy injection contains the entire jet energy, $\sim8\times10^{50}$ ergs.

\begin{figure}
	\includegraphics[width=3.4in]{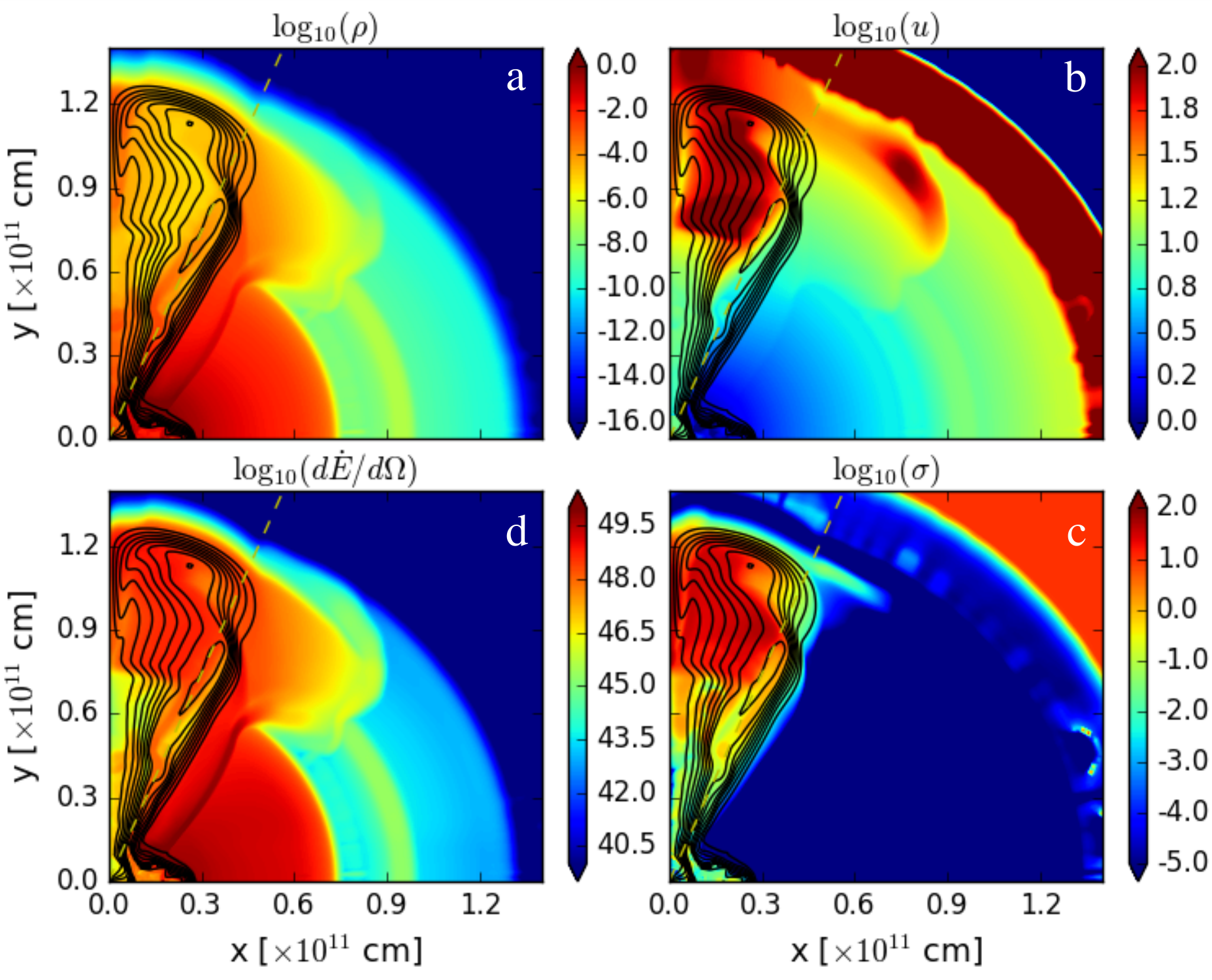}
    \caption{The properties of the jet at breakout, panels show (clockwise) the density (a), the 4-velocity (b), the magnetization (c) and total energy flux per solid angle (d). The black contours depict the poloidal field lines that thread the jet and the inner cocoon. The yellow dashed line marks and opening angle of $20^\circ$, which roughly demarcates the boundary between the jet and the cocoon.}
\label{fig:jet_breakout}
\end{figure}

Figure \ref{fig:jet_breakout} shows the structure of the jet at time $t=4.6$~s from the onset of the jet ($5.1$~s from merger). At this time the jet emerges from the light tail. Going clockwise, the four panels show the density (a), the 4-velocity (b), the magnetization (c) and total energy flux per solid angle (d). The upper jet, which carries the Poynting flux, is shown in panel d in red color. Note that stopping the rotation does not kill the poloidal field. It continues to stretch from the central object to the jet head and maintain enough magnetic pressure to prevent the funnel from collapsing behind the jet. In reality, such an abrupt cut off in the energy injection can occur if the PNS collapses to a BH. In this case the poloidal field lines at the foot of the jet will close uppon themselves and fly out with the jet, leaving a depressurized funnel that will collapse. 

The propagation of the jet through the medium drives a strong forward shock that moves ahead of the jet. As the jet breaks out from the core ejecta and accelerates, the shock accelerates with it and steepens. When the shock reaches the edge of the fast tail, where the optical depth $\tau\sim1$, it breaks out and releases its energy. A similar scenario was dissected by \citet{2012ApJ...747...88N} for the case of hydrodynamic jet breakout in long GRBs. Note that since the shock propagates ahead of the jet in the unmagnetized medium, the physics involved in analyzing it requires only hydrodynamic processes. The jet serves  as a piston, and it does not matter whether it is magnetized or not. 
At this point the shock Lorentz factor is $\Gamma\sim10$ and it is spread over an opening angle of $\sim20^\circ$ (Fig.~\ref{fig:jet_breakout}b). 
Following the analysis of \citet{2012ApJ...747...88N}, we estimate the emitted energy to be of the order of 
\begin{equation}
E_s\sim 6\times10^{46}\left(\frac{r}{2R_\odot}\right)^2~ {\rm ergs}.
\end{equation}
It will reach an on-axis observer approximately {$ t_{obs} = 5.1s - 1.2 \times 10^{11}\rm{cm}/c $} after the merger, and its observed duration will be 
\begin{equation}
\Delta t\simeq 0.02 \left(\frac{r}{2R_\odot}\right)\left(\frac{\theta}{20^\circ}\right)\left(\frac{\Gamma}{10}\right)^{-2}~ {\rm s}.
\end{equation}
Almost the entire breakout emission will be beamed within a $20^\circ$ cone.

After the breakout the jet continues to expand relativistically. It will radiate its energy when reaching the photosphere. For that, however an energy dissipation process must occur, which cannot be modeled in our 2D simulation. Regardless of the dissipation process, due to the high Lorentz factor of the jet material all this energy will be radiated within a cone of $20^\circ$. Thus an observer located at higher latitudes will not be able to see it. 

\subsection{Cocoon breakout}\label{sec:cocoon}
Unlike the relativistic jet, which is confined to a small opening angle {and whose emission is strongly beamed}, the sub-relativistic cocoon expands sideways once it exits the core ejecta, and radiates over much wider angles. 
Therefore, high-latitude emission can {likely} be detected from the jet cocoon. 

The cocoon emerges from the core ejecta alongside the jet, but due to its elongated shape (see Fig.~\ref{fig:jet_core}) the exit at large opening angles is delayed, thus the breakout process from the ejecta is more gradual. Once the cocoon material exits the ejecta into the tail it accelerates under its own pressure.
The inner part of the cocoon, threaded with magnetic field lines   
remains confined to an opening angle of $\lesssim 25^{\circ}$ after the exit, due to the hoop stress of the the magnetic field. The outer cocoon expands to wider angles reaching an opening angle of $\sim 1$ rad. 
A movie showing the emergence process from the ejecta is provided in the
\href{https://youtu.be/Bs5eU_fAv7U}{following link}\footnote{\url{https://youtu.be/Bs5eU_fAv7U}}. 

The cocoon energy, just prior to the breakout, was estimated in eq.~\eqref{eq:Ec} to be $\sim1.3\times10^{50}$ ergs. It's pressure is distributed more or less uniformly in the cocoon. The mass density in the cocoon roughly follows the density distribution of the confining core ejecta, namely it is lower closer to the ejecta's front edge. Therefore the specific enthalpy $h\equiv\frac{\gamma_{add}}{\gamma_{add}-1}\frac{p}{\rho c^2}$ increases towards the cocoon's edge.  When the cocoon exits the core ejecta   
outer shells will accelerate to higher velocities than inner ones, assuming a homologous expansion of the cocoon material and a flat distribution of energy per logarithmic velocity range. This homologous expansion can be seen in Fig.~\ref{fig:jet_breakout}b {through the increase in the expansion velocity and the opening angle of the cocoon}.

\begin{figure}
	\includegraphics[width=\columnwidth]{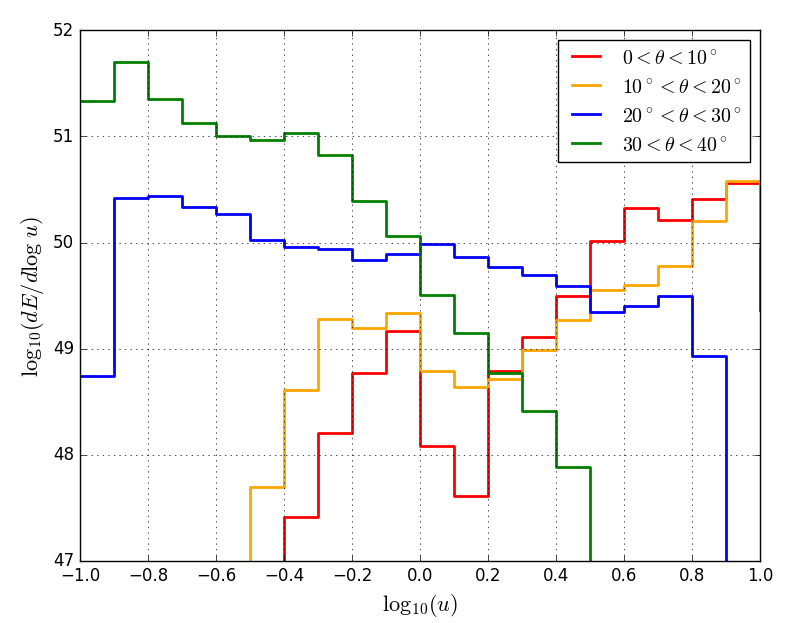}
    \caption{The distribution of energies per logarithmic 4-velocity range ($dE/d\log(u)$) after the breakout from the core ejecta, when the cocoon material reaches a state of homologous expansion. }
    \label{fig:dE_dlogu}
\end{figure}

Figure~\ref{fig:dE_dlogu}, shows the distribution of total energy per logarithmic unit of 4-velocity ($dE/d\log(u)$) after the breakout from the core ejecta, when the cocoon material reaches a state of homologous expansion. The distribution is shown at 4 different angular bins. 
The energy at angles $\theta<20^\circ$ (red and orange lines) is dominated by the jet, thus the energy distribution rises towards high Lorentz factors. At $20<\theta<30$ the energy is dominated by the cocoon and a flat distribution up to $u\simeq5$ is clearly seen. At lower latitudes the acceleration is less efficient, and the energy distribution drops fast at $u\gtrsim0.6$.

The first signature of emission from the cocoon arrives from the breakout phase, as the cocoon's forward shock becomes optically thin. The shock is seen in Fig.~\ref{fig:jet_core} as a sharp drop in the density at the cocoon edge. The breakout emission was analyzed previously for cocoons of long \citep{2012ApJ...747...88N} and short \citep{Gottlieb17b} GRBs in the case of hydrodynamic jets. 
They showed that the emitted power increases strongly with the strength of the shock and the breakout radius. 
To analyze the breakout conditions of the cocoon's shock lets imagine two scenarios: first where only the core ejecta exists without any extended tail, and second where the ejecta is surrounded by a low density fast tail. 

The cocoon breaks out from the ejecta gradually, where cocoon material at larger angles breaks out at later times. Since the ejecta is expanding, a breakout at later times occurs at larger radii. The shock, however  becomes more oblique at larger angles, and thus much weaker (see the \href{https://youtu.be/Bs5eU_fAv7U}{cocoon breakout movie}). 
In our simulation the tip of the cocoon breaks out at a radius $\sim\rdout/(\beta_j-\vdout)\simeq2\times10^{10}$ cm, and the breakout continues up to  $4\times10^{10}$ cm. Without the extended tail the radiation is released to the observer at this stage. The combination of the small radius and the shock obliqueness gives a signal that is too weak to be detected. 

In the case that a fast light material flows ahead of the ejecta, as  the case studied her, the shock breakout is delayed to the radius where it reaches the edge of the fast tail. By that time the shock will have accelerated and became more spherical, thus its shock normal is perpendicular to the breakout surface. Upon breakout its emission will be much stronger and could be detected. Figure \ref{fig:jet_breakout} shows the system during the breakout of the cocoon from the low density tail. The shock is clearly seen at a radius of $\sim1.2\times10^{11}$ cm. It is spread up to an angle of $\sim40^\circ$ and has a spherical shape. Its average Lorentz factor is $\sim 3$.\footnote{The faster moving material seen in panel (b) at $r>1.4\times10^{11}$ cm and the blob at an angle of $40^\circ$ are artifacts of the simulation. They have very low density and bare almost no energy, thus they do not contribute to the emission}

\section{The wide angle $\gamma$-ray emission}\label{sec:radiation}

To calculate the emission from the cocoon's forward shock we apply the method that was developed by \citet{Gottlieb17b} for cocoons of relativistic hydrodynamic sGRB jets. Like in the jet case, since the shock propagates in the unmagnetized ejecta surrounding the cocoon, hydrodynamic methods are sufficient for analyzing its dynamics and calculating the emission. Here we briefly discuss the characteristics of the expected signal and refer to \citet{Gottlieb17b} for more details and a description of the calculation method. The breakout emission is composed of two phases: i) a planar phase that occurs right after the breakout and continues until the shocked gas doubles its radius, and ii) a spherical phase which starts once the planar phase ends and lasts much longer, until the internal energy in the shocked ejecta is radiated away. 
The transition from the planar to the spherical phase can be seen sometimes in the shape of the light curve, but is more prominent in the spectral evolution. The gas just behind the shock is out of thermal equilibrium since not enough photons can be produced in the available time \citep{weaver1976,katz2010,nakar2010,2012ApJ...747...88N}. Further downstream of the shock photons are generated vigorously in the hot gas reducing its temperature and driving it towards thermal equilibrium.  
Therefore there is a general hard to soft evolution with the planar phase being harder than the spherical phase. In addition, during the planar phase, due to light-travel-time angular smoothing, we observe simultaneously emission from regions with different temperatures, resulting in a non-thermal spectrum. In the spherical phase, if the breakout is spherical, the emission at any time is from regions with similar temperatures and thus the spectrum is closer to  a blackbody or Wien spectrum. If the breakout is more oblique, taking place at different times at different angles, the spherical phase emission is also a mix of different temperatures and while the spectrum still shows a hard to soft evolution it remains nonthermal during the entire evolution.
After the transition to the spherical phase the luminosity and temperature drop on a time scale that is comparable to the duration of the planar phase. Therefore, while the emission from the spherical phase continues for a long time, the signal it produces within the $\gamma$-ray window is expected not to be much longer than that of the initial planar phase pulse.   

\begin{figure}
	\includegraphics[width=\columnwidth]{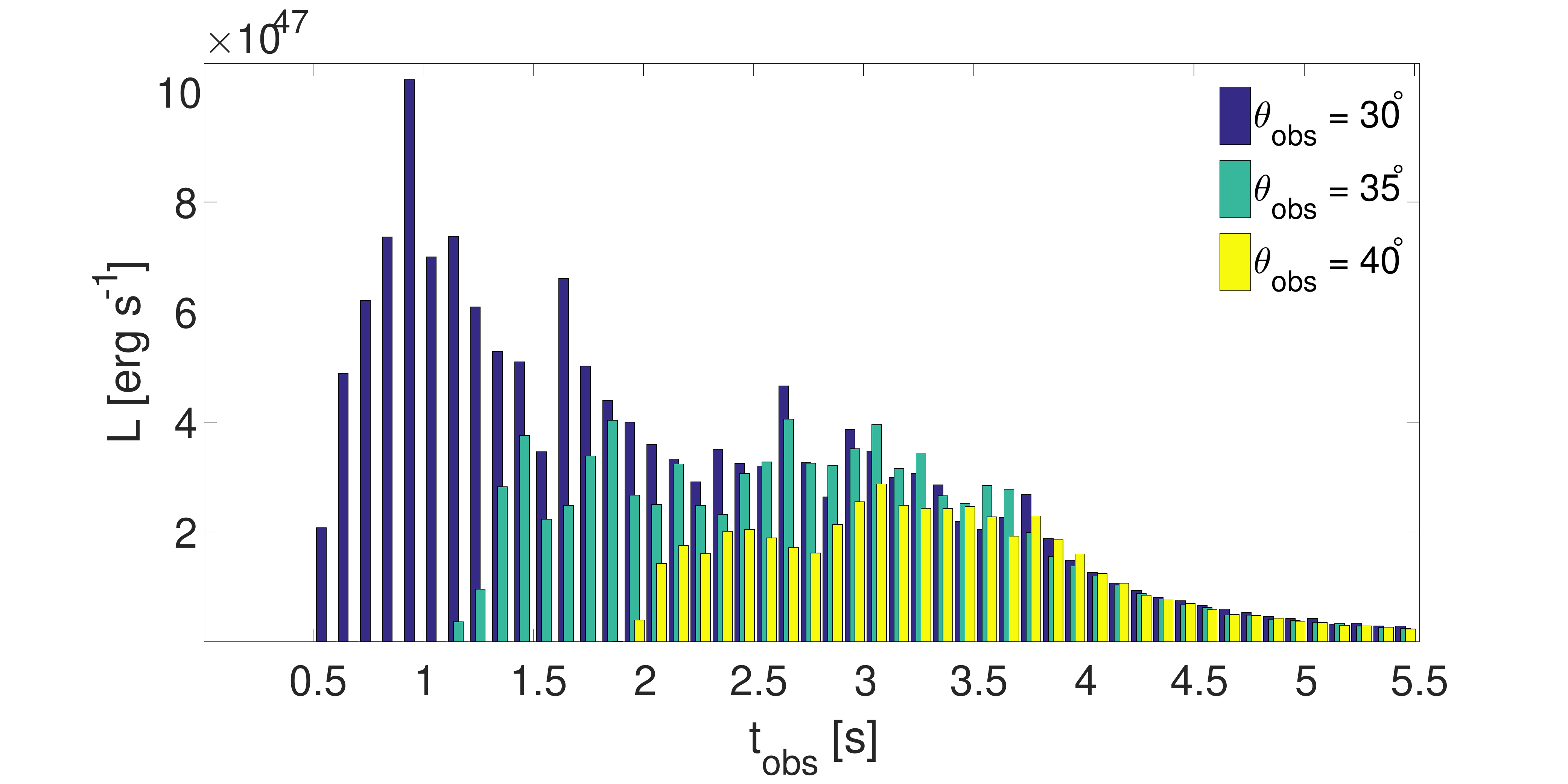}
    \caption{The bolometric luminosity of the shock breakout seeing by viewer at various angles from the jet axis.}
\label{fig:L_c}
\end{figure}

\begin{figure}
	\includegraphics[width=\columnwidth]{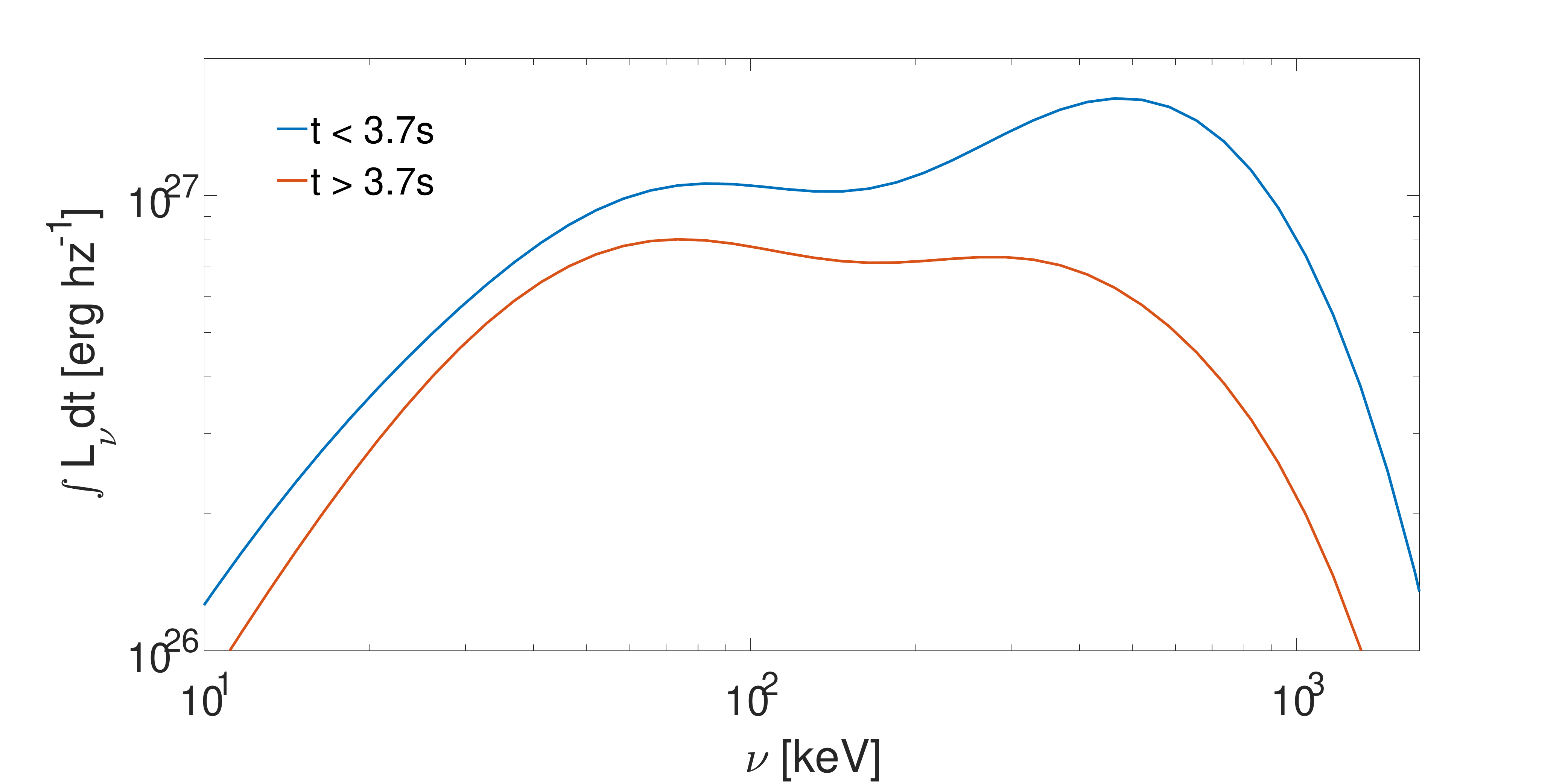}
    \caption{The spectrum of the breakout emission, seen by an observer at $40^\circ$ from the jet axis.}
    \label{fig:spec_c}
\end{figure}

Figure~\ref{fig:L_c} shows the bolometric luminosity that will be observed by viewers at $30^\circ$, $35^\circ$ and $40^\circ$ angles from the jet axis. The time is the observed time from the merger, namely it accounts for the $0.5$~s delay in the formation of the jet. Figure \ref{fig:spec_c}  shows the time-integrated spectrum seen by an observer at $40^\circ$. The spectrum is divided into two epoch. The first epoch (blue) is during the main pulse and the second (red) is during the decay. 
The deviation of the shock breakout from spherical symmetry in our simulation is strong enough so the emission from the two phases is mixed during the entire evolution of the light curve (i.e., planar emission from one angle is mixed with spherical emission from another) and the spectrum shows two peaks both at early and late times. A planar peak around 500 keV and  spherical peak around 60 keV. Still the hard to soft evolution is seen at late times  where  the planar peak becomes less  prominent.  

\section{The $\gamma$-rays counterpart of GW170817}
\label{sec:compare}
On August 17 2017 a GW event was detected by The LIGO Scientific Collaboration and by The Virgo Collaboration, which was attributed to a binary neutron star merger (The LIGO Scientific Collaboration, The Virgo Collaboration 2017d; B. P. Abbott 2017). About $2$~seconds after the GW event Fermi detected a sGRB from a location that is consistent with the localization of the GW signal \citep{Multi,2017GCN.21528....1S}. 

The detected sGRB 170817A was located $40$ Mpc away and had a duration $T_{90}=2\pm0.5$~s that is typical of regular sGRBs. However its energetics and spectral properties are very unusual and distinguish it from other sGRBs. The isotropic equivalent energy $E_{\rm iso}=(5.35\pm1.26)\times10^{46}$ ergs is smaller by three orders of magnitude than the weakest sGRB known. In addition, its light curve consists of two parts with different spectral properties \citep{Multi}. The initial pulse lasts about 0.5s and its spectrum is fitted by a power-law and an exponential cut-off with a peak energy $E_p = 185\pm62$ keV. This pulse is followed by a weaker tail  with a softer spectrum, which is fitted by a black body with $kT \approx 10$ keV. The fluence ratio between the two components is roughly 2:1. This structure is unusual in  sGRBs and, as far as we know, was not observed before.
Analysis by \citep{mansi} show that the observed $\gamma$-rays are most likely produced by a mildly relativistic outflow with $\Gamma\gtrsim2.5$. They also show that no regular scenario that involves a relativistic jet can readily reproduce the observed properties. 

\citep{Gottlieb17b} showed that the observed event can be explained by a jet cocoon breaking out of a mildly relativistic, low density medium and viewed at a large offset ($\sim40^\circ$) from the jet axis. They also found medium parameters that can produce the observables in the context of a hydrodynamic sGRB jet. Here we show qualitatively that the same scenario can account for cocoons of magnetic jets as well. The bolometric lightcurve in Fig.~\ref{fig:L_c} also shows an initial pulse and a tail, and has $T_{90} =2.2 $~s at angles $35^\circ{-}40^\circ$, consistent with observations. Its spectrum also shows hard to soft evolution and two spectral peaks. The agreement with the observations is not perfect though. The $E_{\rm iso}$ we obtain is too high, the spectrum remains non-thermal at all times and the peak energy of the hard component is too high. This is due to the fact that the shock breakout in our case is too strong and not spherical enough. Here we carried out only a single simulation. We did not conduct a search for a setup that produces a signal that matches all the properties of sGRB 170817A, and leave it for a future study. Our work, however, does show that the breakout emission from cocoon shocks is a robust phenomenon, for both hydrodynamic and magnetic jets, and is likely to be detected in more systems in the future. 
Moreover, \citep{Gottlieb17b} showed that the $\gamma$-ray signal depends on the exact system setup and especially on the structure of the fast tail. Based on their scanning of the phase space we expect that there is a setup of the fast tail for which the cocoon breakout from an MHD jet reproduce the observed $\gamma$-rays as well.

\section{Summary}
\label{sec:summary}
In this paper we study the interaction of a relativistic Poynting flux dominated jet with a 
the dynamical mass that was ejected during the merger of two NSs. We use a realistic ejecta mass profile \citep{Hotokezaka2013} with a tail of low mass, fast moving gas.   
The jet is launched in a physical way, by rotating the central compact object, which we represent with a perfectly conducting sphere endowed with a dipole magnetic field. The rotation generates Poynting flux that interacts with the ejecta and collimates into a jet. Our jet has a power of $4\times10^{50}$~erg~s$^{-1}$.

The simulation starts $0.5$~s after the merger. Once formed, the jet propagates through the core ejecta at an average velocity of $v\simeq 0.7 c$ and breaks out into the extended tail it after $\sim 1$~s. This time corresponds to a minimum engine activity time of $t_e=0.3$~s, and is consistent with observational estimations of $t_{\rm e}$ \citep{Moharana2017}.
After the jet breaks out, it accelerates in the tail and assumes a  conical shape with an opening angle of $\sim20^\circ$. It maintains this angle as it continues to propagate further.  

The propagation of the jet through the core ejecta leads to the formation of a hot cocoon around it, which contains the energy expended by the jet for pushing itself through the ejecta. 
At the time of the breakout from the core ejecta the cocoon reaches its maximum energy, $\sim1.3\times10^{50}$ ergs. Once it exits the  ejecta, the cocoon material accelerates homologously and assumes a flat distribution of energy per logarithmic velocity interval. This energy can be tapped to produce the observable radiation, if it undergoes dissipation. 
The existence of the low density fast tail provides such dissipation. It revives the cocoon forward shock, which becomes quite oblique and weak during the breakout from the core ejecta. Propagation through the light tail causes the shock to accelerate to mildly relativistic velocities, under the pressure of the fast propagating cocoon gas. {The shock} breaks out of the low density tail  at a radius of $\gtrsim10^{11}$ cm and emits substantial amounts of radiation that can be observed by distant, off axis observers. 

Before concluding, we note that recently \citet{2017arXiv170807488K} conducted an RMHD study of a jet propagating in a pre-existing funnel, surrounded by a stationary medium. They followed the jet from launching  to the sideways expansion phase after its exit from the funnel.  They found a low-energy component moving at a wide angle from the axis with $\beta\sim0.5$, and identified its source as the outer part of the jet. 
In contrast, in our work all the matter moving at wide angles comes from the inner or the outer cocoon. One possible explanation for the difference is that the jet considered by \cite{2017arXiv170807488K}, which propagates in a pre-existing funnel becomes relativistic very fast and has a very sub-energetic or no cocoon at all. 
We, however, consider a jet interacting with the dynamical ejecta without such a funnel. 
As our jet carves its way through the medium, it forms an energetic cocoon that is not present the other case.  It is thus quite plausible that when the cocoon breaks out alongside the jet in our simulations, its material carries a large enough momentum to prevent the sideways expansion of the jet and cause the large-angle ejecta to be dominated by the cocoon. 

We constructed synthetic lightcurves and spectrum of the breakout emission, seen by viewers at different angles, and compared them with the $\gamma$-ray emission detected from the GW event GW170817. We showed qualitatively that the breakout emission is consistent with the observations. Selecting the parameters of the DME and light tail more carefully can provide valuable information on the physical conditions at the merger site, right after the NSs coalesce and inform us about the physical conditions around these objects.

\section*{Acknowledgements}
We would like to thank K. Hotokezaka for helpful discussions and aid in attaining the ejecta profile used in this work.
{This research was supported by the I-Core center of excellence of the CHE-ISF. OB was supported by the iCore grant. AT was supported by a computational allocation m2401 at the National Energy Research Scientific Computing Center, a DOE Office of Science User Facility supported by the Office of Science of the U.S. Department of Energy under Contract No. DE-AC02-05CH11231. OG and EN were partially supported by an ERC starting grant (GRB/SN) and an ISF grant (1277/13). TP was partially supported by an advanced ERC grant TReX and by a grant from the Templeton foundation.  TP acknowledges kind hospitality at the Flatiron institute while some of this research was done.}




\bibliographystyle{mnras}
\bibliography{mybib,EMGW,another} 







\bsp	
\label{lastpage}
\end{document}